\documentclass[12pt]{article}
\input{epsf}
\usepackage{epsf,cite,epsfig,psfig,float,amssymb,stmaryrd,latexsym}
\usepackage{graphics,psfrag}
\usepackage{a4wide,graphicx}
\usepackage{cite}

\begin{document}

\title{Decay of axial-vector mesons into VP and P$\gamma$}

\author{
L. Roca, J.E. Palomar and E. Oset \\
{\small Departamento de F\'{\i}sica Te\'orica and IFIC,
Centro Mixto Universidad de Valencia-CSIC,} \\
{\small Institutos de
Investigaci\'on de Paterna, Aptd. 22085, 46071 Valencia, Spain}\\
}

\date{\today}

\maketitle  

\begin{abstract} 

We propose a phenomenological Lagrangian for the decay of the
$SU(3)$ nonets of the axial-vector mesons of
$J^{PC}=1^{+-},1^{++}$ into a vector meson and a pseudoscalar
constructed with tensor fields for the vector and axial-vector
mesons. The formulation leads to a good reproduction of the
different decay branching ratios and assuming vector meson
dominance (VMD) it also leads to good results for the radiative
decay of the $a_1$ into pion and photon, and in agreement with
the structure proposed in  the chiral tensor formulation of
radiative decay of axial-vector mesons. The two $SU(3)$
parameters and the mixing angle of  $K_{1A}$ and $K_{1B}$ needed
to give the physical $K_1(1270)$ and $K_1(1400)$ resonances are
also determined. 

 \end{abstract}

\section{Introduction}

The decay of axial-vector mesons of $J^{PC}=1^{+-},1^{++}$
(nonets associated to the $b_1(1235)$ and
$a_1(1260)$ and partners respectively, (see
table~\ref{tab:octets})), into a pseudoscalar (P)  and a vector
meson (V) has received some attention from the perspective of
quark models  \cite{ackleh,bloch} as well as using effective
Lagrangians of the strong interaction 
\cite{wess-zumino,schechter,holstein,bando,meissner,
meissner2,ko-rudaz,li2,li3}. Much of the work have been done in
$SU(2)$ \cite{wess-zumino,bando,meissner2,ko-rudaz,li2} and
general structures have been developed. In particular, an
interesting formulation develops starting from chiral
Lagrangians of pseudoscalar mesons assuming a gauge nature for
the vector fields and introducing them  through covariant
derivatives
\cite{schechter,holstein,bando,meissner,meissner2,ko-rudaz}.
Two main lines of work have developed using the formalisms of
hidden symmetry \cite{bando} and Yang Mills \cite{schechter} and
connections can be done between them \cite{meissner,birse}.
Work has also been done in the $SU(3)$ sector 
\cite{schechter,holstein,meissner,li3}, although it has been
applied mostly to the $SU(2)$ sector, particularly to the study
of the $a_1\to\rho\pi$ decay.

In most of the works the vector field formalism for the vector
mesons is used. There is another way to deal with spin-1
particles assigning them an antisymmetric tensor field
\cite{gasserann,ecker}. The main difference between the tensor
and vector field formalisms stems from the difference in the
vector meson propagator using tensor fields. By introducing
local terms and with basic assumptions of vector meson dominance,
the two formalisms were shown to be equivalent up to
${\cal O}(p^4)$ \cite{eckerPLB} and with only one vector field
(which is not the case studied in the present work). Further
work on the equivalence of the two formalisms, up to local
terms, for general Lagrangians, based on dual transformations of
gauge theories, using path integral techniques, is done in
\cite{bijnens}. The same kind of equivalence exploring the
baryon sector was seen in \cite{borasoy}.

 The tensor formalism has proved
practical in problems of chiral theory in which the vector
mesons are explicitly used in the Lagrangians in order to
study properties of pseudoscalar meson interaction
\cite{ollerND} and pseudoscalar form factors
\cite{oller-pich,palomarFF,portolesFF}, since the vector mesons
are genuine resonances which remain in the large~$N_c$ limit
\cite{ollerND,pelaezNc}, unlike the light scalar mesons which
are dynamically generated by the meson-meson interaction itself
\cite{ollerND,pelaez}. In chiral dynamical approaches, like the
one of \cite{ollerND}, they appear as poles in the driving term
of the interaction.  The use of the tensor formalism in this
case is useful to avoid doublecounting of terms that the lowest
order Lagrangian would already contain should one use the
vector formalism. On the other hand, if one considers the
non-abelian anomaly, the Lagrangian accounting for it, obtained
from the gauged Wess-Zumino term 
\cite{wess-zumino-anomaly,bardeen}, is naturally written in
terms of vector meson fields in the vector representation
although, as suggested in \cite{bijnens}, it could also be
represented in terms of tensor fields. This
anomalous Lagrangian accounts for processes which do not
conserve intrinsic parity, but this is not the case in the
axial-vector decay into vector and pseudoscalar.

The Lagrangian that we use in the present work is also written in
terms of tensor fields. However, for the purpose of evaluating
the $A\to VP$ decays, which is the main aim of the present work,
using the tensor fields, or the analogous $\partial_\mu
V_\nu-\partial_\nu V_\mu$ combination of vector field, is
equivalent since the Lagrangians are only used at tree level.
 We shall see a
practical justification for the use of the 
tensor representation, since
the use of this formalism, together with the vector meson
dominance hypothesis to couple  vector mesons to photons, leads
to a gauge invariant amplitude in the present case which agrees
with the structure and findings for the radiative decays of the
$a_1$ axial-vector meson in the chiral formalism of 
\cite{ecker}. This would not be the case using the straight
vector formalism.

The main motivation for the present study is that evidence is
coming that the $AVP$ vertices play a relevant role in physical
processes like radiative $\phi$ decay \cite{rocaphi} and
$J/\Psi$ decays \cite{uehara,rocaJPsi} through sequential
mechanisms involving the exchange of vector mesons. These
mechanisms were shown to be important in radiative decays of
$\omega$, $\rho$, $\phi$ in \cite{rocaphi,escribano,zaki}. The
works of \cite{rocaphi,uehara,rocaJPsi} came to show that the
sequential mechanism involving the exchange of axial-vector
mesons also play a relevant role in these reactions. With these
problems in mind, it becomes more useful to have at hand a
Lagrangian which combines simplicity, wide applicability and
accuracy. At the same time, most of the work done, for instance
those derived in a chiral formalism
\cite{wess-zumino,schechter,holstein,bando,meissner,
meissner2,ko-rudaz,li2,li3},
pay attention to the $a_1(1260)$ meson, while the $b_1(1235)$
has received much less attention. However, both the $a_1$ and
the $b_1$ mesons and their $SU(3)$ partners appear on an equal
footing in the physical processes which we mentioned before. In
order to construct the needed Lagrangian, satisfying the
mentioned requirements, we will assume $SU(3)$ symmetry and
consider parity and charge conjugation of the particles. One
then obtains two structures involving the commutator and
anticommutator of the  $SU(3)$ matrices associated to the V and
P fields. Although $SU(3)$ breaking terms could also be
present, the results with $SU(3)$ symmetry prove to be accurate
enough within present experimental uncertainties in the data.
$SU(3)$ is broken anyway through the use of the physical
masses. Our Lagrangian contains only one free parameter for the
$a_1$ (and $SU(3)$ partners) and another one for the $b_1$ (and
$SU(3)$ partners) decays. This is in contrast with other models
which already contain three or more parameters for the $a_1$
decay 
\cite{wess-zumino,schechter,holstein,meissner,ko-rudaz,li2,li3}.
In spite of its simplicity we shall see that our Lagrangians
lead to good results for the partial decay widths.

Another issue in our study, implying a new free parameter, is
the mixing angle between the strange members of the nonets, the
$K_{1A}$ and $K_{1B}$, necessary to get the physical
$K_1(1270)$ and $K_1(1400)$ resonances. The experimental fact
that the $K_1(1270)$ decay into $\rho K$  is much larger than 
into $K^*\pi$, and that the $K_1(1400)$ decay into $K^*\pi$ is
much larger than into $\rho K$, is a clear indication that
there is a large mixing between the strange members of the
$SU(3)$ nonets, because $SU(3)$ would imply $K_{1A}$ and
$K_{1B}$ to have a similar amplitude for both decay modes. 
Nevertheless, there is no consensus about the origin of this
mixing angle. In \cite{lipkin} a dynamical origin was
speculated through the coupling of the two $|K_1>$ states to
their decay channels, suggesting the mixing angle to be
$\alpha\sim 45^o$ provided the $K_{1A}$ and $K_{1B}$ states are
degenerate before mixing. Within quark models, a possible
explanation comes from the spin-orbit interaction if $m_q\ne
m_{\overline{q}}$. Concerning the determination of the mixing
angle, most of the works have been based on the fit to the
$K_1$ decays into $\rho K$ and $K^*\pi$ and/or to the ratio 
$B(\tau\to\nu_\tau K_1(1270))/B(\tau\to\nu_\tau K_1(1400))$,
both using symmetry arguments to relate the amplitudes or
either using quark models. Using $K_1$ decay data, the fit of 
\cite{gatto,carnegie} found an angle of $\sim 45^o$, from data
as of 1976, and in \cite{suzuki} the solutions $35^o$, $45^o$
or $57^o$ were found, but the $45^o$ solution was discarded
when considering mass formulas from quark models, which can
relate the masses between the $I=1/2$ members with the mixing
angle. In this latter paper also the $\tau$ decay was
considered to refine the results and the author found that the
experimental data preferred the solution of $35^o$. However,
few years later the experimental results on $\tau$ decay
changed and in the recent work of \cite{cheng} both the $35^o$
and $57^o$ angles were shown to be compatible with the new
data. Using relativized quark models,  the authors of
\cite{blundell} found a large uncertain value of
$-30^o\lesssim\alpha\lesssim 50^o$ from the $\tau$ branching
ratio and $45^o-50^o$ from the $K_1$ decays. In \cite{goldman},
using nonrelativistic constituent quark models to obtain
relations between masses, the authors found
$35^o\lesssim\alpha\lesssim 57^o$. In the work of
\cite{barnes}, the $^3 P_0$ model was used, which is based in
the assumption that the strong decays take place through the
production of a  $q\bar q$ pair with vacuum quantum numbers,
and from the $K_1$ decays a mixing angle $\sim 45^o$ was
obtained. Finally, fitting to charmonium and orthocharmonium
decays, in \cite{suzuki2} a valid range for the mixing angle
between $30^o$ and $60^o$ was obtained. The previous account
summarizes the status of the knowledge on this mixing angle. The
previous works do not use in the fits the data of other $A\to
VP$ decays apart from the $K_1$'s, neither use information of
the dynamics of the amplitude which can be acquired from a
suited Lagrangian. In the present work  we retake the issue, in
view of the accuracy that our Lagrangians provides, and we 
apply our $AVP$ Lagrangian to get the maximum obtainable 
information on the mixing angle from the $A\to VP$ data, and
also the $a_1\to \pi\gamma$ decay. We will use not only the
$K_1$ decay data but also the known available experimental
information of other $A\to VP$ decays which, although do not
depend on the mixing angle, they depend on the couplings of the
Lagrangians and consequently  influence the global fit. The
present study has also the advantage of considering the
dynamics given by our Lagrangians, not only the $SU(3)$
relation between the couplings.

\begin{table}[htpb]
\begin{center}
\begin{tabular}{|c||c|c|c|}\hline 
$J^{PC}$ &$I=1$  &$I=0$ & $I=1/2$ \\ \hline \hline  
 $1^{+-}$  & $b_1(1235)$  & $h_1(1170)$, $h_1(1380)$
    &$K_{1B}$ \\ \hline  
 $1^{++}$  & $a_1(1260)$  & $f_1(1285)$, $f_1(1420)$
    &$K_{1A}$ \\ \hline     
\end{tabular}
\end{center}
\caption{Nonets of axial-vector mesons.}
\label{tab:octets}
\end{table}

\section{The model}

  For the Lagrangians accounting for the $AVP$ vertices, where A symbolizes the
axial-vector mesons, we propose the following expressions

\begin{eqnarray}  \label{eq:LBLA}
{\cal L}_{BVP}&=&\tilde{D} \langle
B_{\mu\nu}\{V^{\mu\nu},P\} \rangle \\
\nonumber
{\cal L}_{AVP}&=&i\tilde{F} \langle A_{\mu\nu}[V^{\mu\nu},P]\rangle 
\end{eqnarray} 

\noindent
where $\langle \rangle$ means $SU(3)$ trace and the
$i$ factor in front of the $\tilde{F}$ is needed
in order ${\cal L}_{AVP}$ to be hermitian.

In Eq.~(\ref{eq:LBLA})
$P$ is the usual $SU(3)$ matrix containing the pseudoscalar
mesons and 
 $B_{\mu \nu}$, $A_{\mu \nu}$, $V_{\mu \nu}$ are 
 $SU(3)$ matrices of the tensor fields 
associated to the axial-vector mesons
 of the $b_1$ and $a_1$ nonet and the
$\rho$ vector meson nonet respectively.
 Assuming the same $SU(3)$ mixture for the 
singlet and
the $I=0$ of the octet that one has for $\phi$
and $\omega$, the vector meson and
axial-vector meson $SU(3)$ matrices $B_{\mu \nu}$, $A_{\mu \nu}$,
$V_{\mu \nu}$ are given by

\begin{eqnarray}
B_{\mu \nu} &\equiv& \left(\begin{array}{ccc} 
 \frac{1}{\sqrt{2}} b^0_1 + \frac{1}{\sqrt{2}}h_1(1170) 
 & b^+_1 & K_{1B}^+\\
b_1^-&- \frac{1}{\sqrt{2}} b^0_1 + \frac{1}{\sqrt{2}}h_1(1170) 
& K^0_{1B}\\
K^-_{1B}& \overline{K}^0_{1B} & h_1(1380)
\end{array}
\right)_{\mu \nu}  \nonumber \\
A_{\mu \nu}&\equiv& \left(\begin{array}{ccc} 
\frac{1}{\sqrt{2}} a^0_1 + \frac{1}{\sqrt{2}}f_1(1285) 
 & a^+_1 & K_{1A}^+\\
a_1^-& - \frac{1}{\sqrt{2}} a^0_1 + \frac{1}{\sqrt{2}}f_1(1285) 
& K^0_{1A}\\
K^-_{1A}& \overline{K}^0_{1A} & f_1(1420)
\end{array}
\right)_{\mu \nu}
\nonumber \\
V_{\mu \nu} &\equiv& \left(\begin{array}{ccc} 
\frac{1}{\sqrt{2}}\rho^0 + \frac{1}{\sqrt{2}} \omega
 & \rho^+ & K^{* +} \\
\rho^-& -\frac{1}{\sqrt{2}} \rho^0
+ \frac{1}{\sqrt{2}} \omega& K^{* 0}\\
K^{* -}& \bar{K}^{*0}& \phi
\end{array}
\right)_{\mu \nu}
\end{eqnarray}

Similarly, the $SU(3)$ matrix, $P$, for the pseudoscalar mesons,
assuming the standard $\eta-\eta'$ mixing \cite{bramon},
 is given by 
 
\begin{eqnarray}
P &\equiv& \left(\begin{array}{ccc} 
 \frac{1}{\sqrt{2}} \pi^0 + \frac{1}{\sqrt{3}}\eta+
 \frac{1}{\sqrt{6}}\eta'
 & \pi^+ & K^+\\
\pi^-& -\frac{1}{\sqrt{2}}\pi^0 + \frac{1}{\sqrt{3}}\eta+
 \frac{1}{\sqrt{6}}\eta'
& K^0\\
K^-& \overline{K}^0 & -\frac{1}{\sqrt{3}}\eta+
 \sqrt{\frac{2}{3}}\eta' 
\end{array}\right).
\end{eqnarray}

In Eq.~(\ref{eq:LBLA}) the fields 
$W_{\mu\nu}\equiv A_{\mu\nu}$, $B_{\mu\nu}$, $V_{\mu\nu}$
 are normalized such that \cite{ecker} 
\begin{equation}
<0|W_{\mu\nu}|W;P,\epsilon>=
\frac{i}{M_W}\left[ P_\mu\,\epsilon_\nu(W)-P_\nu\, 
\epsilon_\mu(W)\right]
\label{eq:tensnorm}
\end{equation}

Eq.~(\ref{eq:tensnorm}) is illustrating because it tells us that
at tree level we would obtain the same results using the
combination $\partial_\mu W_\nu-\partial_\nu W_\mu$ instead of
$W_{\mu\nu}$, with $W_{\mu}$ satisfying the Proca equation, i.e.,
vector formalism for the spin-1 fields.

The structure of the Lagrangians of Eq.~(\ref{eq:LBLA}) is
$SU(3)$ invariant and preserves parity, P, and charge conjugation,
C.

Should we have used vector fields without derivatives
instead of tensor fields in
Eq.~(\ref{eq:LBLA}), the structure that would appear would be the
minimal one, in the sense of number of derivatives,
satisfying the $SU(3)$, parity and charge conjugation symmetries.
For the case of the $1^{++}$ nonet it would be of the form
\begin{equation}
\label{eq:LAvect}
\langle A_\mu[V^\mu,P] \rangle,
\end{equation}
\noindent
which was already derived in \cite{bando} and applied to the
$SU(2)$ sector.
This structure appears also naturally in the
chiral formalism of \cite{schechter,holstein,meissner}
 in addition
to other non-minimal terms 
(see\footnote{There are misprints in the last term of those
equations: a 
commutator symbol in \cite{schechter,meissner} and an extra $\phi$ field in
\cite{meissner} are missing \cite{schechterpriv}.}
Eq.~(3.9) of \cite{schechter} or
(2.59) of \cite{meissner})
\begin{eqnarray}
\langle ( \partial_\mu V_\nu- \partial_\nu
V_\mu)[A^\mu,\partial^\nu\phi]\rangle \nonumber \\
\langle ( \partial_\mu A_\nu- \partial_\nu
A_\mu)[V^\mu,\partial^\nu\phi]\rangle \nonumber \\
\langle ( \partial_\mu A_\nu- \partial_\nu
A_\mu)[( \partial^\mu V^\nu- \partial^\nu
V^\mu),\phi]\rangle .
\label{eq:meissner}
\end{eqnarray}
A particular combination of the first two terms in 
Eq.~(\ref{eq:meissner}) also appears in the chiral formalism of
\cite{meissner2}.
The last term of Eq.~(\ref{eq:meissner}) corresponds to our
Lagrangian of Eq.~(\ref{eq:LBLA}) for tree level calculations, as
we discussed before, given the matrix element of
Eq.~(\ref{eq:tensnorm}). One can prove that the first and second
terms of Eq.~(\ref{eq:meissner}) can be cast in terms of the
third one and the one of  Eq.~(\ref{eq:LAvect}) by using part
integration.
This reduces the freedom in the Lagrangian, at
the level of up to two derivatives, to our tensor structure plus
the vector one of Eq.~(\ref{eq:LAvect}).
Note, however, that the use of derivatives in the present case
does not necessarily mean higher orders in a perturbation
expansion in a small momenta, since derivatives on the vector
fields can produce masses of the massive vectors.
 As mentioned in the
introduction, we shall use only the structure of
Eq.~(\ref{eq:LBLA}), by means of which a good reproduction of
the data is obtained in the $SU(3)$ sector and which naturally
leads, via VMD, to a gauge invariant $A\to V\gamma$ amplitude as
we shall see. We shall also discuss what difference one expects
for the $A\to VP$ decay by using the vector term alone of
Eq.~(\ref{eq:LAvect}). \\

 The width for the decay $A \to VP$ or $B \to VP$ is given by
 
\begin{equation} \label{eq:G1}
\Gamma_{A\to VP}=\frac{q}{8\pi M_A^2}\overline{\sum}|t|^2
\end{equation} 

\noindent
where $q=\frac{1}{2M_A}\lambda^{1/2}(M_A^2,M_V^2,M_P^2)$ is
the momentum of the final particles in the axial-vector rest
frame and the $t$ matrix is given by

\begin{equation}
t_{A\to VP}=\frac{-2\lambda_{AVP}}{M_A M_V}
\left(p'\cdot
p\,\epsilon'\cdot\epsilon
-\epsilon'\cdot p\,\epsilon\cdot p'\right)
\end{equation} 

\noindent
where $p'$,$\epsilon'$, and $p$,$\epsilon$ are the momenta
and polarization vectors of the axial-vector and the
vector mesons respectively. The $\lambda_{AVP}$ are the
coefficient of the $AVP$ vertex obtained using
the Lagrangians of Eq.~(\ref{eq:LBLA}).

The resulting decay width is
\begin{equation}
\Gamma_{A\to VP}=\frac{|\lambda_{AVP}|^2}{2\pi M_A^2}q\left(1+
\frac{2}{3}\frac{q^2}{M_V^2}\right).
\label{eq:GAVP}
\end{equation}
  The coefficients $\lambda_{AVP}$ in Eq.~(\ref{eq:GAVP}) for the
different reactions are given in tables~\ref{tab:coef1} and
\ref{tab:coef2}.
The momentum structure of Eq.~(\ref{eq:GAVP}),
with the weight $2/3$ in the $q^2/M_V^2$
term, is the same as the one obtained in \cite{meissner2}.
Latter on we shall relate the coefficients $\lambda_{AVP}$ to the
pion decay constant, $f$, and will see that our expression
coincides with the one obtained in \cite{meissner2} for the case
$a_1\to\rho\pi$.
\begin{table}[htpb]
\begin{center}
\begin{tabular}{|c | c| c|}\hline 
$b_1^{\pm}\to {K^*} K$& $b_1^0 \to {K^*}^{\pm} K^{\mp}$ & 
$b_1^0 \to {K^*}^{0} K^0$ \\ \hline
$1$ & $\frac{1}{\sqrt{2}}$ & $\frac{-1}{\sqrt{2}}$ \\ \hline \hline
$b_1\to\omega\pi$ &  $b_1\to\rho\eta$& 
$b_1\to\rho\eta'$ \\ \hline
$\sqrt{2}$ &$\frac{2}{\sqrt{3}}$ &$\sqrt{\frac{2}{3}}$ \\ \hline \hline
$a_1^{\pm}\to\rho^{\pm}\pi^0$ &  $a_1^{\pm}\to\rho^0\pi^{\pm}$& 
$a_1^0\to\rho^{\pm}\pi^{\mp}$  \\ \hline
$\pm\sqrt{2}$ &$\mp\sqrt{2}$ &$\mp\sqrt{2}$ \\ \hline \hline
$a_1^{\pm}\to {K^*}^{\pm} \stackrel{(-)}{K^0}$ & $a_1^{\pm}\to \stackrel{(-)}{{K^*}^0}K^{\pm} $ & 
$a_1^0 \to {K^*}^{\pm} K^{\mp}$ \\ \hline
$\mp 1$ & $\pm 1$ & $\mp\frac{1}{\sqrt{2}}$ \\ \hline \hline
$a_1^0 \to {K^*}^0 \overline{K}^0$ & $a_1^0\to{\overline{K}^*}^0 K^0$ &\\ \hline
$\frac{1}{\sqrt{2}}$ & $\frac{-1}{\sqrt{2}}$ &  \\ \hline \hline
$h\to\rho\pi$ & $h\to\omega\eta$ & $h\to\omega\eta'$ \\ \hline
$\sqrt{2}$ & $\frac{2}{\sqrt{3}}$ & $\sqrt{\frac{2}{3}}$ \\ \hline\hline
$h\to K^* K$ && \\ \hline
$\frac{1}{\sqrt{2}}$ & &  \\ \hline\hline
$h'\to K^* K$ & $h'\to\phi\eta$ & $h'\to\phi\eta'$  \\ \hline
$1$ & $\frac{-2}{\sqrt{3}}$ & $2\sqrt{\frac{2}{3}}$  \\ \hline\hline
$f\to {K^*}^{\pm} K^{\mp}$ & $f\to {K^*}^0 \overline{K}^0$ &
$f\to{\overline{K}^*}^0 K^0$ \\ \hline
$\mp\frac{1}{\sqrt{2}}$ & $\frac{-1}{\sqrt{2}}$ & $\frac{1}{\sqrt{2}}$ \\ \hline\hline
$f'\to {K^*}^{\pm} K^{\mp}$ & $f'\to {K^*}^0 \overline{K}^0$ &
$f'\to {\overline{K}^*}^0 K^0$ \\ \hline
$\pm 1$ & $1$ & $-1$ \\ \hline
\end{tabular}
\end{center}
\caption{$\lambda_{AVP}$ for the $b_1$, $a_1$,
 $h\equiv h_1(1170)$, $h'\equiv h_1(1380)$, 
 $f\equiv f_1(1285)$
 and $f'\equiv f_1(1420)$ decays into $VP$. The coefficients
involved in the decay of $b_1$, $h$ and $h'$ 
have to be multiplied by $\tilde{D}$ and those involved in the decay
of $a_1$, $f$ and $f'$ by $i \tilde{F}$.
 The $(-)$ symbol over some $K$ means anti-K when corresponding.}
 \label{tab:coef1}
\end{table}

In the case where there is little phase space
for the decay or it takes place due to the width of the
particles, we fold the expression
for the width with the mass distribution of the particles
as 

\begin{eqnarray} \nonumber
\Gamma_{A\to VP}=\frac{1}{\pi^2}&\int& ds_A ds_V 
Im \left\{\frac{1}{s_A-M_A^2+iM_A\Gamma_A}\right\}
Im \left\{\frac{1}{s_V-M_V^2+iM_V\Gamma_V}\right\}
\cdot \\ 
&\cdot& \Gamma_{AVP}(\sqrt{s_A},\sqrt{s_V})
\Theta(\sqrt{s_A}-\sqrt{s_V}-M_P)
\label{eq:convolution}
\end{eqnarray}
 
\noindent
where $\Theta$ is the step function, 
$\Gamma_{AVP}=\frac{|\lambda_{AVP}|^2}{2\pi s_A}q\left(1+
\frac{2}{3}\frac{q^2}{s_V}\right)$ with
$q=\frac{1}{2\sqrt{s_A}}\lambda^{1/2}(s_A,s_V,M_P^2)$.

 On the other hand for the decay of the $K_1(1270)$ and  $K_1(1400)$ we have to
assume a mixing of the type
\begin{eqnarray} \nonumber 
K_1(1270)&=&\cos(\alpha) K_{1B} -i \sin(\alpha) K_{1A} \\
K_1(1400)&=&\sin(\alpha) K_{1B}+i \cos(\alpha) K_{1A}
\end{eqnarray}

\noindent
 where the prescription is taken to have $\alpha$ comparable to
the definition in 
\cite{gatto,carnegie,suzuki,barnes,blundell}, where no
explicit $i$ factor in the Lagrangian of Eq.~(\ref{eq:LBLA}) 
is considered, which we wrote here explicitly in order to
have a hermitian Lagrangian.

The coefficients in Eq.~(\ref{eq:GAVP}) for the decay of the $K_1(1270)$ and $K_1(1400)$ into
different channels are written in table~\ref{tab:coef2}.

\begin{table}[htpb]
\begin{center}
\begin{tabular}{|c|c|c|c|}\hline 
$K_1^{\pm}\to \rho^0 K^{\pm}$& $K_1^{\pm}\to \rho^{\pm} \stackrel{(-)}{K^0}$ & 
$\stackrel{(-)}{K_1^0} \to \rho^{\mp} K^{\pm}$ & 
$\stackrel{(-)}{K_1^0} \to \rho^0 \stackrel{(-)}{K^0}$\\ \hline & & &  \\[-0.45cm]
$\pm\frac{1}{\sqrt{2}}(c{\tilde{D}}-s{\tilde{F}})$ & $\pm (c{\tilde{D}}-s{\tilde{F}})$ & $\pm (c{\tilde{D}}-s{\tilde{F}})$ &
$(+)-\frac{1}{\sqrt{2}}(c{\tilde{D}}-s{\tilde{F}})$ \\ \hline \hline
$K_1^{\pm}\to K^{\pm}\omega$ & $K_1^0\to K^0\omega$ & $\overline{K}_1^0\to \overline{K^0}\omega$& \\ \hline & & &  \\[-0.45cm]
$\pm\frac{1}{\sqrt{2}}(c{\tilde{D}}-s{\tilde{F}})$ & $\frac{1}{\sqrt{2}}(c{\tilde{D}}-s{\tilde{F}})$ &
$\frac{-1}{\sqrt{2}}(c{\tilde{D}}-s{\tilde{F}})$&\\ \hline\hline
$K_1^{\pm}\to K^{\pm}\phi$ & $K_1^0\to K^0\phi$ &
$\overline{K}_1^0\to \overline{K}^0\phi$& \\ \hline & & &  \\[-0.45cm]
$\pm(c{\tilde{D}}+s{\tilde{F}})$ & $c{\tilde{D}}+s{\tilde{F}}$ & $-(c{\tilde{D}}+s{\tilde{F}})$&\\ \hline\hline
$K_1^{\pm}\to {K^*}^{\pm}\pi^0$& $K_1^{\pm}\to \stackrel{(-)}{{K^*}^0}\pi^{\pm}$ & 
$\stackrel{(-)}{{K_1}^0} \to {K^*}^{\pm} \pi^{\mp}$ & 
$\stackrel{(-)}{{K_1}^0} \to \stackrel{(-)}{{K^*}^0}\pi^0$\\ \hline & & &  \\[-0.45cm]
$\pm\frac{1}{\sqrt{2}}(c{\tilde{D}}+s{\tilde{F}})$ & $\pm (c{\tilde{D}}+s{\tilde{F}})$ & $\pm (c{\tilde{D}}+s{\tilde{F}})$ &
$(+)-\frac{1}{\sqrt{2}}(c{\tilde{D}}+s{\tilde{F}})$ \\ \hline \hline
${K_1}^{\pm} \to {K^*}^{\pm} \eta$ & 
${K_1}^{0} \to {K^*}^{0} \eta$ & $\overline{K}_1^0 \to{\overline{K}^*}^{0} \eta$&\\ \hline & & &  \\[-0.45cm]
$\pm\frac{2}{\sqrt{3}}s{\tilde{F}}$ & $\frac{2}{\sqrt{3}}s{\tilde{F}}$ & $\frac{-2}{\sqrt{3}}s{\tilde{F}}$ &
 \\ \hline \hline
${K_1}^{\pm} \to {K^*}^{\pm} \eta'$ & 
${K_1}^{0} \to {K^*}^{0} \eta'$ & $\overline{K}_1^{0} \to
{\overline{K}^*}^{0} \eta'$&\\ \hline
$\pm\sqrt{\frac{3}{2}}(c{\tilde{D}}-\frac{1}{3}s{\tilde{F}})$ & $\sqrt{\frac{3}{2}}(c{\tilde{D}}-\frac{1}{3}s{\tilde{F}})$ &
 $-\sqrt{\frac{3}{2}}(c{\tilde{D}}-\frac{1}{3}s{\tilde{F}})$ &\\ \hline
\end{tabular}
\end{center}
\caption{$\lambda_{AVP}$ for the decay of $K_1(1270)$ into
$PV$. In the coefficients, $c\equiv\cos(\alpha)$ and
$s\equiv\sin(\alpha)$. The coefficients for the decays of the
$K_1(1400)$ are the same but changing $F\to -F$ and
interchanging $s\leftrightarrow c$.}
 \label{tab:coef2}
\end{table}

\section{Results and discussion}

Considering the values given in the particle data table
\cite{pdg} for
the decay widths (see table~\ref{tab:decays}),
we carry out a best fit to these data to obtain the ${\tilde{D}}$,
${\tilde{F}}$ and $\alpha$ parameters.

\begin{table}[htpb]
\begin{center}
\begin{tabular}{|c|c|c|}\hline 
Reaction & BR \% & partial width (MeV) \\ \hline\hline
 $K_1(1270)\to \rho K$ & $42\pm 6$ & $38\pm 10$ \\ \hline
 $K_1(1270)\to K^* \pi$ & $16\pm 5$ &$14\pm 6$ \\ \hline
 $K_1(1270)\to \omega K$ & $11\pm 2$& $10\pm 3$\\ \hline
 $K_1(1400)\to K^* \pi$ & $94\pm6$ & $164\pm 16$ \\ \hline
 $K_1(1400)\to \rho K$ & $3\pm3$ & $5\pm 5$\\ \hline
 $K_1(1400)\to \omega K$ & $1\pm 1$ & $1.7\pm 1.7$\\ \hline
 $b_1\to\omega\pi$ & $75\pm 25$ & $110\pm 40$\\ \hline
 $f'\to K^* K$ & $50\pm 25$ & $28\pm 14$\\ \hline
 $h\to \rho \pi$ & $50\pm 50$ & $180\pm 180$\\ \hline
 $a_1\to \rho \pi$ & $50\pm 50$ &  $210\pm 210$\\ \hline
 $h'\to K^* K$ & $50\pm 50$ & $45\pm 45$\\ \hline
\end{tabular}
\end{center}
\caption{Reactions considered in the fit to obtain the $D$,
$F$ and $\alpha$ parameters. The data for the $K_1$ decays
are taken from \cite{pdg}. For the other reactions there are
no explicit data but we can infer from there an approximate
value with a large uncertainty.}
\label{tab:decays}
\end{table}

In table~\ref{tab:decays} we show the decays for which there
are data for the branching ratios with their errors. For the
$K_1$ decays, the ratios taken are as in the PDG \cite{pdg}. In
the case when the decay has been seen but no numbers are provided
in the PDG, we assume the value $(50\pm 50)$\% for the branching
ratio, implying that in the fit we put as experimental input
that the partial decay width is smaller than the total width.
These data will generally weigh little in the fit but including
them prevents solutions with partial decay widths for some
channels unreasonably larger than the total width. In the case
where the PDG gives only a partial decay width as "dominant" with
no number, ($b_1\to \omega\pi$), we have
taken for the fit a branching ratio of $(75\pm 25)$\% which allows
any value from more than half to total. In the case where there
are two channels presented as "dominant", ($f'\to K^*K$),
we have taken  $(50\pm 25)$\% branching ratio for
each of these channels.

 We observe that the  $K_1(1270)$ meson decays largely into
$K \rho$ with the  $K^* \pi$ channel suppressed, while the
$K_1(1400)$ decays largely into  $K^* \pi$ with the $K \rho$
channel suppressed.  This feature is what demands the large
mixing between the $K_{1A}$ and $K_{1B}$.
We also see that the
$a_1$ decays into $\rho \pi$, with  the $\omega \pi$
decay mode forbidden, while the $b_1$ decays into  $\omega
\pi$. 

As we can see in table~\ref{tab:decays}, the decays of
$K_1(1270)$ and $K_1(1400)$ have good experimental data. But
the theoretical expression for these decays manifests 
certain symmetries under some transformation of the
parameters: First of all the theoretical value for the decay
is exactly the same if one interchanges ${\tilde{D}}$ and
 ${\tilde{F}}$ and
changes  $\alpha\to\pi/2-\alpha$. The other symmetry is
manifested if one makes the following substitutions:
$\tilde{D}\to-\sin(\alpha)\sqrt{\tilde{D}^2+\tilde{F}^2}$, 
$\tilde{F}\to\cos(\alpha)\sqrt{\tilde{D}^2+\tilde{F}^2}$ and
$\alpha\to\arctan\left(\frac{-\tilde{D}}{\tilde{F}}\right)$.
These two symmetries\footnote{Apart from these symmetries,
typical only for the $K_1$ decays, there is another symmetry
in all the $A\to PV$ decays in the global sign of $\tilde{D}$
and $\tilde{F}$ but with a fixed value for
$\tilde{D}\tilde{F}$ if the mixing angle $\alpha$ is
restricted to be between  $0$ and $90$ degrees.}  lead to
four different set of parameters as mathematically equivalent
solutions if one only considers the $K_1(1270)$ and
$K_1(1400)$ decays, which are shown
 in table \ref{tab:resultsfit1}.

\begin{table}[htpb]
\begin{center}
\begin{tabular}{|c|c|c|c|}\hline & & &  \\[-0.45cm]
${\tilde{D}}$ (MeV) & ${\tilde{F}}$ (MeV)& $\alpha$ (degrees) &
$\chi^2/dof$\\ \hline
$-1250\pm 80$ & $1400\pm 100$ & $62\pm 3$ & $1.394$ \\ \hline
$-1650\pm 100$ & $880\pm 90$ & $42\pm 3$ & $1.394$ \\ \hline
$-1400\pm 100$ & $1250\pm 80$ & $28\pm 3$ & $1.394$ \\ \hline
$-880\pm 80$ & $1650\pm 100$ & $48\pm 3$ & $1.394$ \\ \hline
\end{tabular}
\end{center}
\caption{Results of the fit to the data including only the
$K_1(1270)$ and $K_1(1400)$ decays. ($dof$ is the number of
degrees of freedom $dof=(\# exp. data)-(\# parameters))$.}
\label{tab:resultsfit1}
\end{table}

These solutions are similar to those found in  \cite{suzuki}.
 These symmetries in the solutions of
the fit can be broken if one introduces other decays which do
not depend on $\alpha$. The problem is that there are very few
of these data in the PDG, only from  $a_1\to K^* K$ one could
infer a reasonable fair value for the branching ratio to be used
in the fit. But this reaction is told in the PDG to be
controversial and then we do not use it in the fit.  

We next  include also in the fit the last five channels of
table~\ref{tab:decays},  corresponding to "seen" or "dominant"
in the PDG, with prescription for the branching ratios and
errors explained above. The results  for the parameters
obtained from the fit are shown in
table~\ref{tab:resultsfit2}. 

\begin{table}[htpb]
\begin{center}
\begin{tabular}{|c|c|c|c|c|}\hline & & &  \\[-0.45cm]
&${\tilde{D}}$ (MeV) & ${\tilde{F}}$ (MeV)& $\alpha$ (degrees) &
$\chi^2/dof$\\ \hline
$(1)$&$-1240\pm 80$ & $1380\pm 100$ & $62\pm 3$ & $0.687$ \\ \hline
$(2)$&$-1330\pm 90$ & $1250\pm 80$ & $29\pm 3 $ & $0.746$ \\ \hline
$(3)$&$-980\pm 100$ & $1520\pm 100$ & $47\pm 4$ & $1.207$ \\ \hline
\end{tabular}
\end{center}
\caption{Results of the fit including the data of
 table~\ref{tab:decays}}
\label{tab:resultsfit2}
\end{table}

We can see that the results are very similar to those found in
table~\ref{tab:resultsfit1} because the uncertainties of the new
data included are very large and weigh very little in the fit.
The main novelty of the new fit is that the solution of
$\alpha=42$ degrees in table~\ref{tab:resultsfit1}  disappears
and, as seen in  table~\ref{tab:resultsfit2}, it seems to prefer
the solutions $(1)$ and $(2)$ to the one of $\alpha=47$ degrees.
This conclusion would be the same one obtained in \cite{suzuki}
for different reasons, since the extra decays evaluated here
were not considered in \cite{suzuki}. In any case we should note
that the precise value of the $\chi^2$ function in
table~\ref{tab:resultsfit2} is tied to the way the unknown data
have been entered in the fit (see table~\ref{tab:decays}) and
that, in any case, a fit with a $\chi^2/dof$ close to $1$ is an
acceptable solution.

We now comment on the differences of using the Lagrangian of 
Eq.~(\ref{eq:LAvect}) instead of the one of Eq.~(\ref{eq:LBLA})
used so far.
In the vector formalism, see Eq.~(\ref{eq:LAvect}), the
amplitude and width are given by 

\begin{equation}
t=-\lambda_{AVP}\,\epsilon'\cdot\epsilon\qquad;\qquad
 \Gamma_{A\to VP}=\frac{|\lambda_{AVP}|^2}{8\pi M_A^2}
q\left(1+\frac{1}{3}\frac{q^2}{M_V^2}\right),
\label{eq:tvector}
\end{equation}

\noindent
where now $\tilde{D}$ and $\tilde{F}$ would have a different
normalization. The difference of Eq.~(\ref{eq:tvector}) with 
Eq.~(\ref{eq:GAVP}) is a factor $2$ in the $q^2$ term in the
bracket. However, this term is reasonable small in all the
decay channels and thus the numerical differences in the decay
rates between the two formalisms are very small. Later on we
shall nevertheless show that using the tensor formalism leads
naturally to a gauge invariant amplitude for the radiative decay of the
$a_1$ resonance, which is not the case if one uses the Lagrangian
of Eq.~(\ref{eq:LAvect}).\\

\begin{figure}
\centerline{\protect\hbox{
\psfig{file=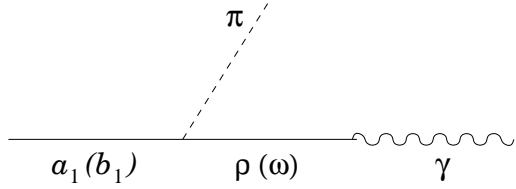,width=0.45\textwidth,silent=}}}
\caption{Diagram for the $a_1$ and $b_1$ decay into
$\pi\gamma$ through vector meson dominance.}
\label{fig:figure}
\end{figure}

   Next we pass to study the radiative decay of the $a_1$ and
$b_1$ into  $\pi \gamma$.  We assume vector meson dominance
 and hence the mechanism for the decay is
represented by the Feynman diagram of Fig.~\ref{fig:figure}.

 The    case of
the $a_1$ decay proceeds with the exchange of a $\rho$ meson
while the decay of the $b_1$ requires the exchange of the
$\omega$ meson.  In addition to the Lagrangians which we
have, we need the vector meson-photon coupling which is given
in \cite{ecker} in the tensor formalism by

\begin{equation}
{\cal L}_{V\gamma}=-e\frac{F_V}{2}\lambda_{V\gamma} V_{\mu\nu}^0
(\partial^\mu A^\nu - \partial^\nu A^\mu)
\label{eq:LVMD}
\end{equation}

\noindent
with $\lambda_{V\gamma}=1,\frac{1}{3},-\frac{\sqrt{2}}{3}$
for  $V^0=\rho^0,\omega,\phi$ respectively.

Furthermore one also needs the vector meson propagator in the
tensor formalism which is also given in \cite{ecker} as

\begin{eqnarray} \label{eq:prop_4indices}
&<&0|T\{W_{\mu\nu}W_{\rho\sigma}\}|0>=i{\cal
D}_{\mu\nu\rho\sigma} =\\ \nonumber
&=&i\frac{M_W^{-2}}{M_W^2-P^2-i\epsilon}\left[g_{\mu\rho}\,
g_{\nu\sigma}\,(M_W^2-P^2)
+g_{\mu\rho}\,
P_\nu \, P_\sigma-g_{\mu\sigma}\,P_\nu \, P_\rho-(\mu \leftrightarrow \nu) \right]
\end{eqnarray}

In view of this, the decay width is given by
Eq.~(\ref{eq:G1}), where $t$ is now given by

\begin{equation}
t=\frac{2\lambda_{AVP}\lambda_{V\gamma} F_Ve}{M_A M_V^2}
\left(p'\cdot
p\,\epsilon'\cdot\epsilon
-\epsilon'\cdot p\,\epsilon\cdot p'\right)
\label{eq:tAVP}
\end{equation}

\noindent 
which exhibits explicitly gauge invariance. Should we have used
the vector couplings of Eq.~(\ref{eq:LAvect}) for the $AVP$ plus
the vector formalism for the vector-photon coupling of
\cite{bramon}, we would have obtained only the term
$\epsilon'\cdot\epsilon$ which does not fulfill gauge invariance
and is unacceptable to represent the process, something that was
already pointed out in  \cite{brown}. On the other hand,
should we have use Eqs.~(\ref{eq:LBLA}) and (\ref{eq:LVMD}),
replacing $W_{\mu\nu}$ by $\partial_\mu W_\nu - \partial_\nu
W_\mu$ and using the standard propagator for 
$\partial_\mu W_\nu - \partial_\nu W_\mu$, (the same as
Eq.~(\ref{eq:prop_4indices}) except for the contact terms
proportional to $(M_W^2-P^2)$), we would have got zero,
indicating that one would have to add contact terms in the vector
formalism to make it equivalent to the tensor one, in the line of
the claims made in other works trying to show the equivalence
between the vector and tensor formalisms
\cite{eckerPLB,borasoy,bijnens,birse}.

  Using the couplings
obtained before for the $AVP$ vertices plus those of the  
$V\gamma$ vertex, we
find for the radiative decay width of the $a_1$ resonance of
$380\pm 50\textrm{ KeV}$, $320\pm 40\textrm{ KeV}$ or $470
\pm 60\textrm{ KeV}$ for the solutions $(1)$, $(2)$ and $(3)$
of table~\ref{tab:resultsfit2} respectively, in reasonable
good agreement  with the experimental value $640\pm
246\textrm{ KeV}$. This is not the case of the $b_1$ decay
for which we get a width $32\pm 4\textrm{ KeV}$,
$36\pm5\textrm{ KeV}$ or $19 \pm 4\textrm{ KeV}$ for the
solutions $(1)$, $(2)$ and $(3)$, which is too small compared
with experiment, $230 \pm 60\textrm{ KeV}$. The reason for
this drastic reduction from the   $a_1$ to the $b_1$ is the
factor $1/3$ of the $\omega \gamma$ coupling with respect to
the $\rho \gamma$ coupling. Since for this reason the vector
meson dominance term is so much reduced with respect to the
case of the $a_1$ decay, it is not surprising that other
mechanisms can also account for this decay channel. However,
the $b_1\to\pi\gamma$ decay width is still much smaller than
that of the $a_1$, around a factor 3.  Such possible
mechanisms would thus be smaller than the vector meson
dominance mechanisms for the case of the $a_1$ decay, in
which  case the decay width provided by the vector meson
dominance mechanism should be relatively accurate as it is the
case, accounting for more than $2/3$ of the total
$a_1\to\pi\gamma$ decay width.

  It is interesting to establish comparison of the result
obtained here for the radiative decay width of the $a_1$
resonance with those of \cite{ecker}. This comparison can be
carried out at the analytical level by recalling the
structure of the Lagrangian in \cite{ecker}

\begin{equation} \label{eq:LAf}
{\cal L}=\frac{F_V}{2\sqrt{2}} 
\langle A_{\mu\nu}f_{-}^{\mu\nu}\rangle
\end{equation}

\noindent
where $f_{-}^{\mu\nu}$ is defined in \cite{ecker} and
provides, in our case,
the pseudoscalars and photon fields.
From the Lagrangian of Eq.~(\ref{eq:LAf}) one can derive the
amplitude for the $a_1^+\to\pi^+\gamma$ decay obtaining

\begin{equation}
t=-\frac{iF_Ae}{f M_A}
\left(p'\cdot
p\,\epsilon'\cdot\epsilon
-\epsilon'\cdot p\,\epsilon\cdot p'\right)
\end{equation}

\noindent where $f=92.4\textrm{ MeV}$ and $e$ is taken
positive. This latter amplitude  can be compared to the one
in Eq.~(\ref{eq:tAVP}) using that for the $a_1^+\to\pi^+\gamma$
decay $\lambda_{V\gamma}=1$ and $\lambda_{AVP}=-\sqrt{2}iF$
obtaining that

\begin{equation}
\tilde{F}=\frac{1}{2\sqrt{2}}\frac{F_A}{F_V}\frac{M_V^2}{f}
\label{eq:FvsFA}
\end{equation}

It is interesting to recall that the $F_A$ parameter is
related, through the Weinberg sum rule \cite{weinberg},
to the $F_V$ and $f$ parameters by 
$F_A=\sqrt{F_V^2-f^2}$,
and using values of $F_V$ and $f$ from \cite{ecker},
then $F_A \simeq 123\textrm{ MeV}$. Using this value
in Eq.~(\ref{eq:FvsFA}) one obtains 
${\tilde{F}}\simeq 1800\textrm{ MeV}$, 
to be compared with the values obtained
in table~\ref{tab:resultsfit2}.
It is also interesting to recall that $F_A$ is related to the
$L_{10}$ coefficient of the second order meson 
chiral Lagrangian \cite{chiPT} through 
$L_{10}^{V+A}=-\frac{F_V^2}{4M_V^2}+\frac{F_A^2}{4M_A^2}$.
With the value obtained in our fit for $\tilde{F}$, and
consequently for $F_A$, the same qualitative agreement
obtained in \cite{ecker} with the empirical value of $L_{10}$
holds also here. This allows to relate the parameter
 $\tilde{F}$ of our effective Lagrangian (Eq.~\ref{eq:LBLA})
with the $L_{10}$ parameter of the chiral Lagrangians.

Apart from the comparison of the coefficients at the numerical
level, it is also illustrative to compare the analytical
expression of the coefficient of Eq.~(\ref{eq:FvsFA}) with the
corresponding one used in \cite{meissner2}. By using the vector
meson dominance values of \cite{ecker}, $F_V=\sqrt{2}f$, $F_A=f$,
$M_A=\sqrt{2}M_V$, we obtain $\tilde{F}=M_V^2/4f$, and
considering the $a_1^+\to\rho^+\pi^0+\rho^0\pi^+$ and the
coefficients of $\lambda_{AVP}$ in Table~\ref{tab:coef1}, we
obtain, for the coefficient of Eq.~(\ref{eq:GAVP}),
\begin{equation}
\frac{|\lambda_{a_1^+\rho^+\pi^0}|^2
+|\lambda_{a_1^+\rho^0\pi^+}|^2}{2\pi M_A^2}
=\frac{M_\rho^2}{16\pi f^2}
\end{equation}
which coincides with Eq.~(2.22) of \cite{meissner2}. \\

We can also perform another study including the experimental
$a_1\to\pi\gamma$ decay width into the fit, in which case we
obtain:

\begin{table}[htpb]
\begin{center}
\begin{tabular}{|c|c|c|c|c|}\hline & & &  \\[-0.45cm]
&${\tilde{D}}$ (MeV) & ${\tilde{F}}$ (MeV)& $\alpha$ (degrees) &
$\chi^2/dof$\\ \hline
$(1)$&$-1230\pm 80$ & $1380\pm 90$ & $62\pm 3$ & $0.743$ \\ \hline
$(2)$&$-1320\pm 90$ & $1270\pm 80$ & $29\pm 3 $ & $0.859$ \\ \hline
$(3)$&$-960\pm 90$ & $1540\pm 100$ & $47\pm 4$ & $1.129$ \\ \hline
\end{tabular}
\end{center}
\caption{Results of the fit including the data of 
table~\ref{tab:decays}
and the $a_1\to\pi\gamma$ decay.}
\label{tab:resultsfit3}
\end{table}
 
With this final set of parameters we can predict the
widths of all the $A\to VP$ decay widths not included
 in the particle data table. The results for the three possible
 solutions of table~\ref{tab:resultsfit3} are shown in
  table~\ref{tab:predictions}. We make predictions for all possible decays,
many of them yet unobserved. The errors quoted are statistical, but
large uncertainties should be assumed in cases where the phase
space is only allowed due to the tails of the resonance mass
distributions through Eq.~\ref{eq:convolution}. It is however
instructive to see that all widths predicted are well within the
values of the total width of the decaying particle,
 which we have also written in the table for comparison.
\begin{table}[h]
\begin{center}
\begin{tabular}{|c|c|c|c|c|c|}\hline & & &  \\[-0.45cm]
Reaction & $\Gamma_{\textrm{tot}}^{\textrm{exp}}\textrm{ (MeV)}$ 
&$\Gamma_i^{\textrm{exp}}\textrm{ (MeV)}$ &
$\Gamma_i\textrm{ (MeV)}$ $(1)$ & $\Gamma_i\textrm{ (MeV)}$ $(2)$
&$\Gamma_i\textrm{ (MeV)}$ $(3)$ \\ \hline\hline
 $a_1\to\pi\gamma$ & $425\pm 175$ & $0.64\pm 0.25$ &$0.37\pm 0.05$  & $0.31\pm0.04 $ &  $0.46\pm 0.06 $     \\ \hline
 $K_1(1270)\to \rho K$& $90\pm 20$ & $38\pm 10$&$48\pm 5$    & $46\pm 5$    &  $47\pm 5 $ \\ \hline
 $K_1(1270)\to K^* \pi$ & & $14\pm 6$& $10\pm 4$ &$8\pm 4$      &  $6\pm 4$ \\ \hline
 $K_1(1270)\to \omega K$ & & $10\pm 3$& $12.8\pm 1.3$& $12.4\pm 1.3$&  $12.6\pm 1.4$ \\ \hline
 $K_1(1400)\to K^* \pi$ & $174\pm 13$&  $164\pm 16$&$143\pm14$  & $145\pm 14$  &  $146\pm 17$ \\ \hline
 $K_1(1400)\to \rho K$ & & $5\pm 5$& $6\pm4$ & $7\pm 4$     &  $4\pm 4 $ \\ \hline
 $K_1(1400)\to \omega K$& &  $1.7\pm 1.7$&$2.1\pm 1.2$ & $2.4\pm 1.3$&  $1.3\pm 1.2$ \\ \hline
 $b_1\to\omega\pi$& $142\pm 9$ &  $110\pm 40$&$114\pm 14$      & $130\pm 18$  &  $69\pm 14$ \\ \hline
 $f'\to K^* K$ & $55.5\pm 2.9$ &  $28\pm 14$&$42\pm 5$ 	      & $34\pm 4$    &  $51\pm 6$ \\ \hline
 $h\to \rho \pi$ &  $360\pm 40$& $180\pm 180$& $290\pm 40$       & $330\pm 50$  &  $180\pm 30$ \\ \hline
 $a_1\to \rho \pi$& &  $210\pm 210$& $260\pm 30$  &  $220\pm 30$ &  $320\pm 40$ \\ \hline
 $h'\to K^* K$  &  $91\pm 30$  & $45\pm 45$&  $42\pm 5$    & $48\pm 7$    &  $25\pm 5 $ \\ \hline
 $a_1\to K^* K$ &  & & $31 \pm 4$        &  $26\pm 3$   &  $39\pm 5$  \\ \hline
$b_1\to K^* K$   & & &$9.1 \pm 1.1$     &  $10.5\pm 1.4$ &$5.6\pm 1.1$  \\ \hline
$b_1\to\rho\eta$ & & &$16 \pm 2 $       &  $18\pm 3$   &  $9.8\pm 1.9$  \\ \hline
$b_1\to\rho\eta'$& & &$0.81 \pm 0.10$   &  $0.93\pm 0.13$  &  $0.49\pm 0.10$  \\ \hline
$h\to\omega\eta$ & & &$17 \pm 2$        &  $19\pm 3$    &  $10\pm 2$  \\ \hline
$h\to\omega\eta'$&  & &$2.5 \pm 0.3$    &  $2.9\pm 0.4$ &  $1.6\pm 0.3$  \\ \hline
$h\to K^*K$      & & &$17.7 \pm 2$      &  $20\pm 3$    &  $11\pm 2$  \\ \hline
$h'\to \phi\eta$ & & &$2.2 \pm 0.3$     &  $2.6\pm 0.4$ &  $1.4\pm 0.3$  \\ \hline
$h'\to\phi\eta'$ & & &$0.43 \pm 0.05$   &  $0.49\pm 0.07$  &  $0.26\pm 0.05$  \\ \hline
$f\to K^*K$     & $24.0\pm 1.2$ & & $3.8 \pm 0.5$     &  $3.2\pm 0.4$ &  $4.8\pm 0.6$  \\ \hline
\end{tabular}
\end{center}
\caption{Partial widths obtained for all the $A\to VP$ decays with the three different
possible solutions of the $\tilde{D}$, $\tilde{F}$
 and $\alpha$ parameters.
The theoretical errors shown are only due to the uncertainties in these
parameters.}
\label{tab:predictions}
\end{table}
  \\
  
Recently, a paper dealing with $\tau \to \pi\pi\pi \nu_\tau$ has
appeared \cite{portoles} in which five different structures to
account for the $AVP$ coupling in the tensor formalism have been
derived. Although the structures derived there are formally
different to the one we propose, it is easy to see that at tree
level the terms $O^4$ and $O^5$ of Eq.~(12) of \cite{portoles}
give an identical structure for the amplitude
to that of our Lagrangian and so would do a linear combination of
$O^2$ and $O^3$ . The $O^1$, however, breaks explicitly
$SU(3)$ symmetry and hence has no room in our $SU(3)$ symmetric
approach. In practical terms, the results of the present work
indicate that should one use the formalism of \cite{portoles} at
tree level to deal with $A\to VP$ decay, one would obtain very
good agreement with the data by taking, for instance,
 the $O^4$ term alone.
 This,
of course, affects only the octet of the $a_1$, not the $b_1$
which we have also studied here.
 
\section{Summary}
 
We have addressed the decay of an axial-vector into a vector
and a pseudoscalar meson looking for a Lagrangian which can
reproduce all existing data while making predictions for all
yet  unobserved allowed channels involving all the particles of
the $SU(3)$ nonets. We found that a basic Lagrangian involving
commutator and anticommutator of the fields, using the tensor
representation for the vector and axial-vector mesons, was
rather accurate and, at the same time, simple enough to be used
in intermediate steps of more complicated hadronic processes.

We have also shown that the combination of our $AVP$ Lagrangian
with Vector Meson Dominance leads to an amplitude for the
radiative decay of the $a_1$ into $\pi \gamma$, which formally
agrees with the one obtained in the chiral formulation of vector
meson and axial couplings, relating the $\tilde{F}$ parameter of
the $AVP$ coupling to the $F_V$ parameter of the $V\gamma$
coupling in VMD and to the coefficients of the meson chiral
Lagrangians. The tensor formalism produces small numerical
differences in the predictions for the different decays of the
axial-vector mesons with respect to the vector formalism without
derivatives, yet it leads naturally to a gauge invariant
amplitude for the radiative decay of the $a_1$ resonance while
this vector formalism leads to a non invariant one. From the
studied strong decays we found three acceptable solutions for
the parameters and the mixing angle of the strange
axial-vectors, with two of them with angles around $30$ and $60$
degrees, slightly favored with respect the solution around $45$
degrees. This is the maximum information that can be obtained
from all the $A\to VP$ present decay data. The present
determination of the mixing angle has the advantage from
previous works of using all the $A\to VP$ available decay data,
not only $K_1$ decays, and of considering the dynamics given by
a suitable Lagrangian. Regarding the prediction for unobserved
channels, it is interesting to observe that all the predicted
decay rates are well within the boundaries of the total decay
widths. Since the predictions of the three different solutions
for some channels are quite different, the measurement of some
of them would be most welcome in order to find out the actual
mixing and the value of the coupling parameters.

The simple form derived for the Lagrangian has made easier the
implementation  of mechanisms involving exchange of axial-vector
mesons which contribute in processes of radiative decays of
$\phi$ and $J/\Psi$ which had not been discussed till recently.
With tests of hadronic models and particularly chiral dynamics
been conducted in physical processes occurring at higher
energies, an increasing attention will have to be payed to the
role of axial-vector mesons. The work in the present paper makes
this task easier.

\section*{Acknowledgments}
We would like to acknowledge useful discussions with
J.~Portol\'es.
Two of us, J.E.P. and L.R., acknowledge support from the
Ministerio de Educaci\'on y Ciencia. 
This work is
partly supported by DGICYT contract number BFM2003-00856,
and the E.U. EURIDICE network contract no. HPRN-CT-2002-00311.

\end{document}